\documentclass[letter,10pt,twocolumn]{article}
\usepackage{float}
\usepackage{graphicx}
\usepackage{dcolumn}
\usepackage{amsmath}
\usepackage[dvips]{epsfig}
\usepackage{subfigure}

\begin{document}


\title{Enhanced single-photon emission from a quantum dot in a micropost microcavity}

\author{Jelena Vu\v{c}kovi\'{c}, David Fattal, Charles Santori, Glenn S. Solomon and\\
Yoshihisa Yamamoto\\
Quantum Entanglement Project, ICORP, JST\\
Ginzton Laboratory, Stanford University, Stanford CA 94305}

\maketitle

\begin{abstract}

We demonstrate a single-photon source based on a quantum dot in a micropost microcavity that exhibits a large Purcell factor together with a small multi-photon
probability. For a quantum dot on resonance with the cavity, the spontaneous emission rate is increased by a factor of five, while the probability to emit two or
more photons in the same pulse is reduced to 2\% compared to a Poisson-distributed source of the same intensity. In addition to the small multi-photon
probability, such a strong Purcell effect is important in a single-photon source for improving the photon outcoupling efficiency and the single-photon generation
rate, and for bringing the emitted photon pulses closer to the Fourier transform limit.\\ \\
\end{abstract}

Generation of single photons at a well defined timing or clock is crucial for practical implementation of quantum key distribution (QKD) \cite{ref:BB84}, as well
as for quantum computation \cite{ref:qucomp_prop1} and networking based on photonic qubits \cite{ref:Cirac,ref:Duan}. Three different criteria are taken into
account when evaluating the quality of a single-photon source: high efficiency, small multi-photon probability (measured by the second-order coherence function
$g^{(2)}(0)$), and quantum indistinguishability. For example, high efficiency and small $g^{(2)}(0)$ are required, but quantum indistinguishability is not
necessary for BB84 QKD \cite{Edo}. On the other hand, for almost all other applications in quantum information systems we need photons that are indistinguishable
and thus produce multi-photon interference.

One of the popular approaches to generation of single photons is based on a pulsed excitation of a quantum dot (QD) combined with spectral filtering
\cite{ref:Santori01,ref:Michler00,ref:Gerard01,ref:Zwiller01}. Although a single quantum dot by itself can be used to generate single photons
\cite{ref:Santori01}, the efficiency of such a system is poor, as the majority of emitted photons are lost in the substrate. In addition, the radiative lifetime
can be as long as 1 ns, which is greater than the dephasing time estimated to be $\sim$0.9 ns at 4K \cite{ref:Bayer}. Emitted photons are unlikely to be
indistinguishable, with coherence lengths shorter than the radiative limit (Fourier transform limit). Finally, the single-photon generation rate is low, as
determined by the long excitonic lifetime. Microcavities can help in correcting all of these deficiencies \cite{ref:Gerard98,ref:Gerard01,ref:Michler00}. The
radiative lifetime of an emitter on resonance with the cavity can be decreased significantly below the dephasing time, bringing the emitted photon pulses closer
to the Fourier transform limit. Moreover, the spontaneous emission rate can be enhanced, and a large fraction of spontaneously emitted photons can be    coupled
into a single cavity mode, thereby increasing the outcoupling efficiency. By employing a QD in a micropost microcavity, our group has recently demonstrated
efficiencies close to 40\% \cite{ref:Matt_prl}. We have also proved that consecutive photons emitted from such a source are largely indistinguishable, with a
mean wave-packet overlap as large as 0.81 \cite{ref:Charlie}. In this article, we demonstrate an improved single-photon source based on a quantum dot in a
micropost microcavity that exhibits a large Purcell factor ($F_p$=5) together with a small multi-photon probability ($g^2(0)$=2\%).

\begin{figure}[ht]
\begin{center}
\subfigure{\epsfig{figure=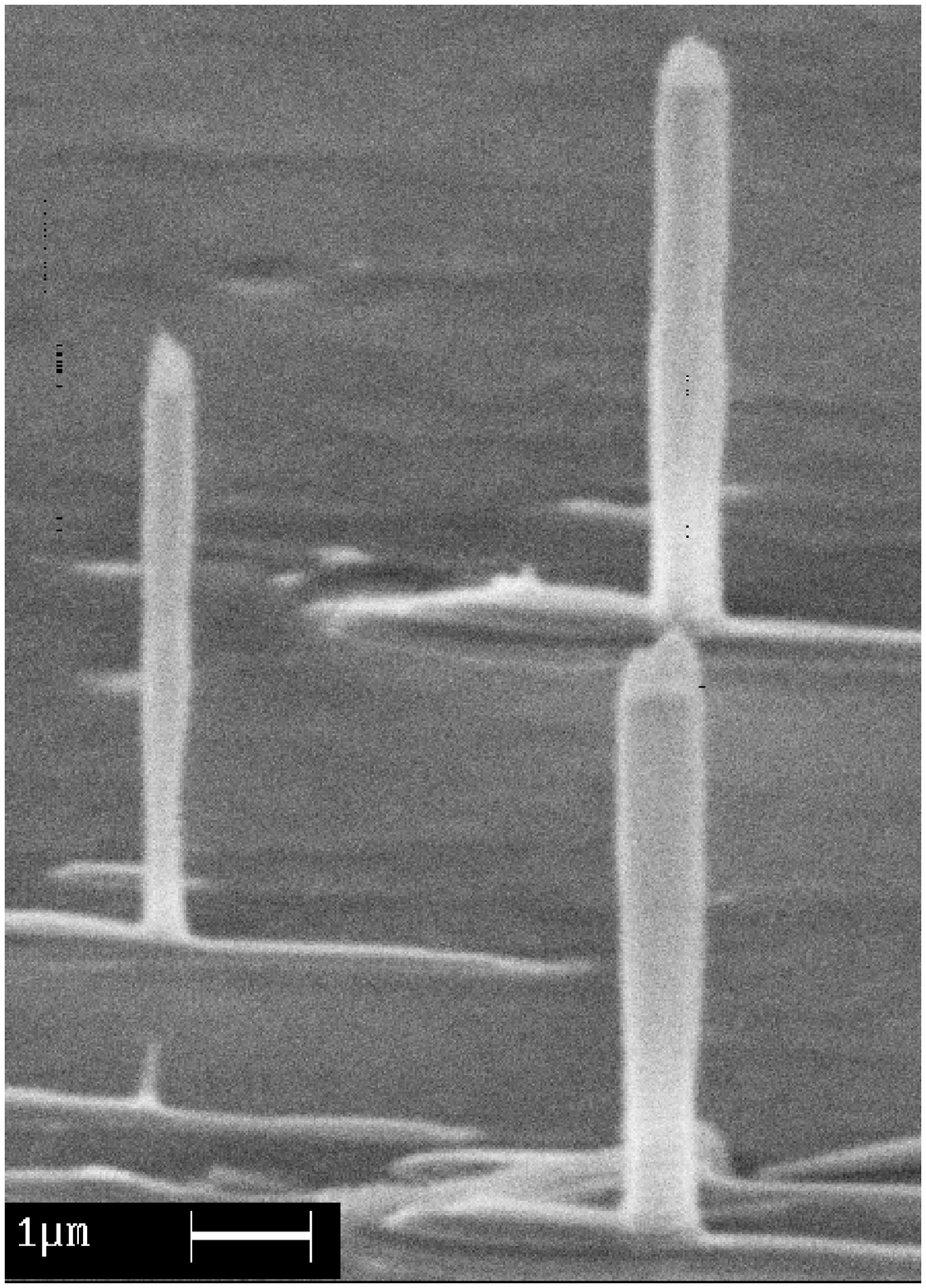,width=0.45\linewidth}} \subfigure{\epsfig{figure=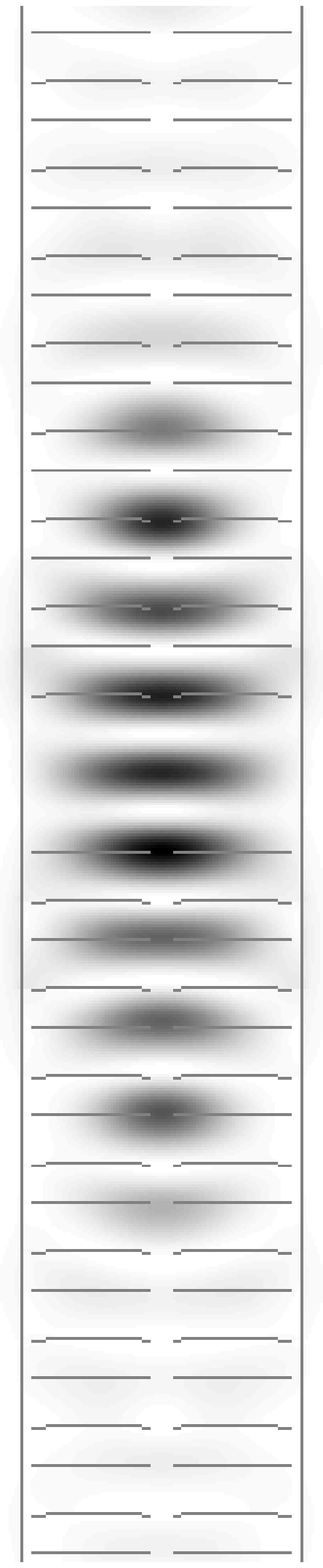,width=0.3\linewidth}} \caption{Left: Scanning electron micrograph
showing a fabricated array of microposts. Remaining sapphire dust (used as etch mask) is visible at the top of the structures. Right: Simulated electric field
intensity of the fundamental HE$_{11}$ mode in a cavity.} \label{fig:fabs}
\end{center}
\end{figure}

Our single-photon source consists of a self-assembled InAs QD embedded in the middle of a GaAs spacer in a distributed Bragg reflector (DBR) micropost
microcavity. The GaAs spacer is approximately one optical wavelength thick (274 nm), and sandwiched between twelve DBR mirror pairs on top, and thirty DBR mirror
pairs on bottom. Each DBR pair consists of a 68.6 nm thick GaAs and a 81.4 nm thick AlAs layer (both layers are approximately quarter-wavelength thick). The
advantages of microposts relative to other microcavities are that the light escapes normal to the sample, in a single-lobed Gaussian-like pattern, and that it is
relatively straightforward to isolate a single QD in them. The micropost structures (shown in Fig. \ref{fig:fabs}) were constructed by a combination of
molecular-beam epitaxy (MBE) and chemically assisted ion beam etching (CAIBE). MBE was used to grow a wafer consisting of DBR mirrors and a GaAs spacer with
embedded self-assembled QDs. Microposts with diameters ranging from 0.3 $\mu$m to 5 $\mu$m and heights of 5 $\mu$m were fabricated in a random distribution by
CAIBE, with Ar$^+$ ions and Cl$_2$ gas, and using sapphire (Al$_2$O$_3$) dust particles as etch masks. Sapphire was chosen as a mask because of its chemical
stability and hardness, which enabled larger etch depths than other mask-materials. Unfortunately, these same properties impede its removal from the tops of the
structures, without endangering the microposts \cite{ref:etch_sapphire}. The presence of sapphire on top of the structures decreases their quality factors and
consequently reduces outcoupling efficiencies. This can be confirmed theoretically by the Finite Difference Time Domain (FDTD) calculation of the Q-factor of the
fundamental HE$_{11}$ mode (whose field pattern is also shown in Fig. \ref{fig:fabs}) in a micropost with and without sapphire on top. The details of the FDTD
method can be found in our earlier publication \cite{ref:YYpaper2}; the FDTD unit cell size used in this case is 5 nm. We assume that the DBR layers of the
simulated post have the same parameters as in the experimentally studied structures, that the the post has perfectly straight walls, diameter of 0.5 $\mu$m, and
that refractive indices of GaAs, AlAs, and Al$_2$O$_3$ are 3.5, 2.9, and 1.75, respectively. Without sapphire, the Q-factor of the HE$_{11}$ mode is 2600 and its
wavelength is 882 nm; with 0.5 $\mu$m thick sapphire disk on top of the post (with diameter equal to that of the post), the Q-factor drops to 1400, while the
mode wavelength remains unchanged. Due to the irregular shapes of the fabricated posts, the HE$_{11}$ mode is typically polarization-nondegenerate, and many
microposts have only one or two QDs on resonance with this fundamental mode.

In our experimental setup, the sample with microposts is placed in a liquid He cryostat. The microposts are excited from a steep angle by Ti:sapphire laser
pulses, 3 ps in duration, with a 76 MHz repetition rate, and resonant with higher-level confined states of a QD. With resonant excitation, the favored absorption
of a single electron-hole pair is expected, and no carriers are created in the vicinity of the QD, suppressing emission wavelength drift due to charge
fluctuations. The emission from the QD is collected normal to the sample surface, and then directed towards a streak camera preceded by a spectrometer, for
time-resolved photoluminescence measurements. The spectral resolution of the system is 0.1 nm, together with a time resolution of 25 ps. For photon correlation
measurements, the collected emission is first spectrally filtered with bandwidth of 0.1 nm and then directed towards a Hanbury Brown and Twiss-type (HBT) setup.
In the HBT setup, photon counters are placed at both outputs of a non-polarizing beamsplitter for detection. The electronic signals from the counters are sent to
a time-to-amplitude converter followed by a multi-channel analyzer computer card, which generates a histogram of the relative delay time $\tau=t_2-t_1$ between a
photon detection at one counter ($t_2$) and the other ($t_1$). For the detailed description of the setup, please refer to \cite{ref:Santori01}.

\begin{figure}[ht]
\begin{center}
\epsfig{figure=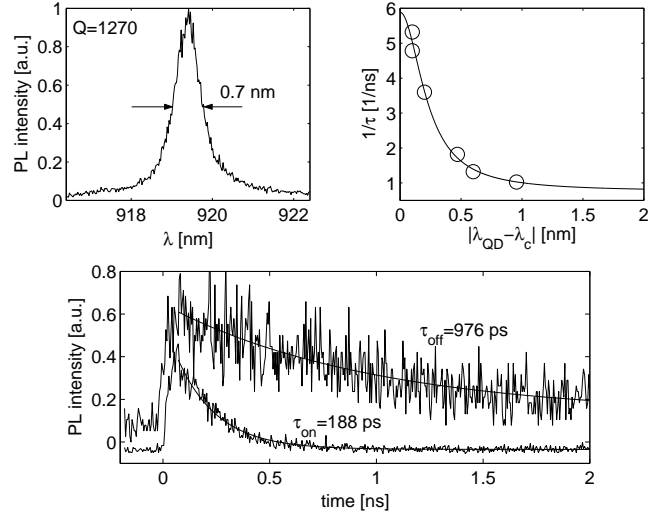,width=3.375in} \caption{Top-left: Background emission filtered by the cavity.
Top-right: Decay rate of the emission line as a function of the absolute value of its detuning from the cavity resonance ($|\lambda_{QD}-\lambda_c|$). The dot
emission wavelength was tuned by changing the sample temperature within the 6 K - 40 K range. Bottom: time-dependent photoluminescence from the emission line
on-resonance with the cavity, as opposed to this same emission line off-resonance.}
\label{fig:lifetimes}
\end{center}
\end{figure}

We performed lifetime and $g^{(2)}$ measurements on a QD chosen for its bright emission under resonant excitation. Tuning of the sample temperature was used to
tune the emission wavelength relative to the cavity resonance \cite{ref:Kiraz01}. In this particular case, the dot emission wavelength was tuned away from the
cavity resonance by increasing the sample temperature from 6 K to 40 K \cite{ref:other_dots}. Fig. \ref{fig:lifetimes} (bottom) shows the time-resolved
photoluminescence of an emission line on- and off-resonance with the cavity mode. The decay lifetime differs by a factor of five for these two cases. The decay
rate for this emission line as a function of the absolute value of its detuning from the cavity resonance ($|\lambda_{QD}-\lambda_c|$) is plotted with circles in
the top-right plot of Fig. \ref{fig:lifetimes}. The solid line corresponds to the Lorentzian fit to the experimental data \cite{ref:Gerard98}, and a good match
is observed between our experiment and the theoretically predicted behavior. The fitting parameters are the linewidth, maximum and minimum of the Lorentzian,
while it is assumed that its central wavelength is equal to $\lambda_c$, the cavity resonance wavelength. $\lambda_c$ and the quality factor ($Q=$1270) are
determined from the photoluminescence intensity taken at high pump powers (see the top-left plot of Fig. \ref{fig:lifetimes}). The cavity resonance wavelength
red-shifts by roughly 0.3 nm with increasing temperature in the studied range, and this shift is included in plotting the data. Up to fivefold spontaneous
emission rate enhancement (Purcell factor $F_p$) is observed for the dot coupled to a cavity, as opposed to the dot off-resonance (top-right plot of Figure
\ref{fig:lifetimes}).

The theoretical limit of the Q-factor of the studied microposts is 4000, which is the value calculated for a planar cavity (before etching) without any
absorption losses and inaccuracies in the growth of DBR layers. The maximum Q-factor of a micropost with a finite diameter has to be below this limit, due to the
additional loss mechanisms in the transverse directions \cite{ref:YYpaper2} and the presence of sapphire dust particles, as discussed above. According to the
detailed theoretical treatment of the QD micropost device \cite{ref:YYpaper2,ref:Pelton02}, the cavity Q-factor and Purcell factor can be much larger
($Q\sim$10000 and $F_p\sim$100) for optimized cavity designs with 15 and 30 DBR pairs on top and bottom, respectively, perfectly straight cavity walls, and a QD
located at the cavity center.

\begin{figure}[ht]
\begin{center}
\epsfig{figure=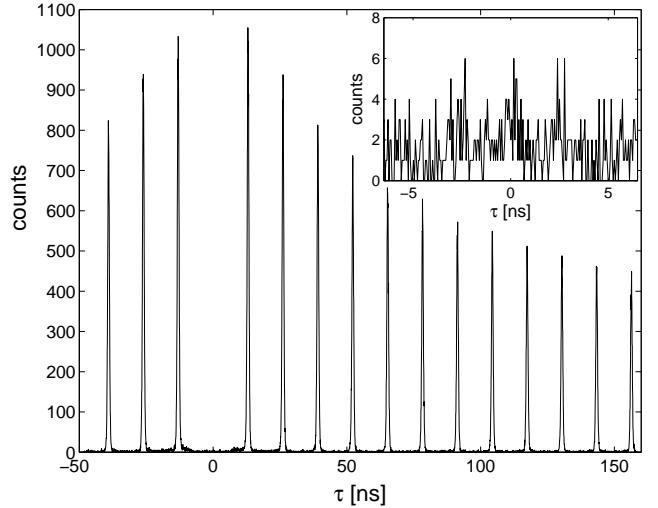,width=3.375in}\caption{Photon correlation histogram for a QD on resonance with the cavity (the lifetime of this dot is shown in Fig.
\ref{fig:lifetimes}), under pulsed, resonant excitation. The inset depicts the magnified central portion (from $\tau$=-6.5 ns to 6.5 ns) of the histogram. The
missing central peak (at $\tau$=0) indicates a large suppression of multi-photon pulses.} \label{fig:g2}
\end{center}
\end{figure}

A photon-correlation measurement for this same dot on resonance with the cavity is shown in Fig. \ref{fig:g2}. The histogram is generated using the described HBT
setup. The distance between peaks is 13 ns, corresponding to the repetition period of pulses from the Ti:sapphire laser. The decrease in the height of the side
peaks as $|\tau|$ increases indicates the dot blinking behavior generally observed under resonant excitation, and can be approximated with a double-sided
exponential \cite{ref:Santori01}. The vanishing central peak (at $\tau$=0) indicates a strong antibunching and a large suppression of multi-photon pulses. The
probability of generating two and more photons for the same laser pulse compared to a Poisson-distributed source of the same intensity ($g^2(0)$) is estimated
from the ratio of the areas of the central peak and the peaks at $|\tau|\rightarrow\infty$. Each area is calculated by integrating all the counts within the
integration window centered at the peak and without subtracting any background counts. The area of the peak at $|\tau|\rightarrow\infty$ is estimated from the
decaying exponential fit to the heights of the side peaks. $g^2(0)$ is estimated to be equal to 2\% for an integration window of 4 ns.  The integration window is
chosen so that the contribution of the peak tails outside it to the peak area can be neglected (it is below 1\%). The width of the histogram peaks is determined
by the photon counter timing resolution (0.3 ns) and the excitonic lifetime \cite{ref:Santori01}. Owing to the strong Purcell effect (excitonic lifetime below
0.2 ns), this small multiphoton probability should be preserved even for the repetition period much smaller than 13 ns (e.g., 2 ns). If the integration window is
reduced to 1 ns, $g^2(0)$ drops to 1\%. Depending on the application of the photon-source, a different definition of the two-photon probability may be necessary.
For example, in the interference experiment with two consecutive photons emitted from a dot, the relevant parameter is the probability to emit two photons in the
same pulse, as opposed to the probability to emit one photon in each of the two consecutive pulses \cite{ref:Charlie}. This parameter, which we denote as $g$, is
calculated from the ratio of the area of the central peak to the area of the nearest side peak, and is equal to 0.9\% for the integration window of 4 ns. The
difference between $g$ and $g^2(0)$ is a result of the blinking behavior of the dot.

In summary, we have studied the effect of a microcavity on single photons emitted from a QD, and have demonstrated an improved single-photon source that exhibits
a large Purcell factor ($F_p$=5) together with a small multi-photon probability ($g^2(0)$=2\%). In addition to a small multi-photon probability, such a strong
Purcell effect is important in a single-photon source for improving the photon outcoupling efficiency and the single-photon generation rate, and for bringing
the emitted photon pulses closer to the Fourier transform limit.\\

\begin{bf}Acknowledgement \end{bf}\\\\
This work is partially supported by MURI UCLA/0160-G-BC575. The authors would like to thank A. Scherer and T. Yoshie from Caltech for providing access to CAIBE
and for helping with fabrication.\\
Jelena Vu\v{c}kovi\'{c} is with the Department of Electrical Engineering, Stanford University, Stanford, CA 94305. E-mail: jela@stanford.edu. Charles Santori is
also at the IIS, University of Tokyo, Tokyo, Japan. Glenn S. Solomon is also at the SSPL, Stanford University, Stanford, CA 94305. Yoshihisa Yamamoto is also at
NTT Basic Research Labs, Atsugishi, Japan.

\bibliographystyle{unsrt}
\bibliography{APL03b_Vuckovic_ref}

\begin{thebibliography}{10}
\bibitem{ref:other_dots}
This QD is almost exactly on resonance with the cavity at low temperature, so
  by heating the sample and increasing the QD emission wavelength, one also
  increases the detuning from the cavity resonance and the radiative lifetime.
  The opposite process is observed if the QD emission wavelength is initially
  smaller than the cavity resonance. In addition to a good correspondence
  between the cavity resonance linewidth and the Lorentzian linewidth, these
  processes indicate that the cavity QED has a dominant effect on the radiative
  lifetime.
\end{thebibliography}

\end{document}